\documentclass[12pt]{iopart}
\usepackage{graphicx}

\begin{document}


\title[Topological defects and hybrid inflation in light of three-year
  WMAP observations]{Limits on defects formation and hybrid
  inflationary models with three-year WMAP observations}

\author{Aur\'elien A Fraisse}
\address{Princeton University Observatory, Peyton Hall, Princeton, NJ
  08544, USA}
\ead{fraisse@astro.princeton.edu}

\begin{abstract}
  We confront the predicted effects of hybrid inflationary models on
  the Cosmic Microwave Background (CMB) with three years of 
  \emph{Wilkinson Microwave Anisotropy Probe} (WMAP) observations.
  Using model selection, we compare the ability of a simple flat
  power-law $\Lambda$CDM model to describe the data to that of hybrid
  inflationary models involving global or local cosmic strings, or
  global textures.  We find that it is statistically impossible to
  distinguish between these models:  they all give a similar
  description of the data, the maximum ratio of the various Bayesian
  evidences involved being never higher than
  $\mathrm{e}^{0.1 \pm 0.5}$.  We then derive the maximum contribution
  that topological defects can make to the CMB, and place an upper
  bound on the possible value of cosmic strings' tension of 
  $G\mu \leq 2.1 \times 10^{-7}$ (68\%~confidence limit).  Finally, we
  give the corresponding constraints on the D-term strings' mass
  scale, as well as limits on the F- and D-term coupling constants
  ($\kappa$ and $\lambda$) and inflationary scales
  ($M$~and~$\sqrt{\xi}$).
\end{abstract}


\maketitle


\section{Introduction}

Spontaneous symmetry-breaking (SSB) phase transitions were first
invoked to explain the evolution of condensed matter systems, such as
ferromagnets, as a function of their
temperature~\cite{Kittel96,Feynman89}.  It was soon realized that what
was true for classical quantum objects and was one of the backbones of
the theories of superfluidity and
superconductivity~\cite{VilenkinShellard}, could also be applied in
the context of quantum field theory.

Kirzhnits~\cite{Kirzhnits72} and Kirzhnits and
  Linde~\cite{KirzhnitsLinde72} first exhibited the analogy between
  elementary particles' sym\-metries and these classical systems,
  arguing that vacuum symmetries that are broken today can be restored
  at high temperatures in the early Universe.  This idea, later
  developed at the same time by Kirzhnits and
  Linde~\cite{Kirzhnits74}, Weinberg~\cite{Weinberg74} and Dolan and
  Jackiw~\cite{Dolan74}, led to the concept of a grand unified group
  of symmetries $G$ that could have broken down into the group of
  symmetries describing the forces of Nature we know today, namely 
  $SU(3) \times U(1)_\mathrm{em}$~\cite{Pati73,Georgi74}.  The
  previous success of Glashow~\cite{Glashow61} and
  Weinberg~\cite{Weinberg67} in describing the electromagnetic and
  weak forces in the context of the single group of symmetry
  $SU(2) \times U(1)$ gave even more weight to the idea of grand
  unification.

However, when a spontaneous sym\-metry-breaking phase tran\-si\-tion
oc\-curs betwe\-en a given group of symmetry and another one
exhibiting a non-trivial set of degenerate ground states, it is
expected that topological defects will form at the border of domains
which end up in different minima.  Kibble~\cite{Kibble76,Kibble80}
proposed a mechanism explaining the formation of defects as a function
of the homotopy groups of the manifold of degenerate vacua, as well as
a qualitative description of their cosmological evolutions.

It was soon realized that the production of monopoles, point-like
defects, as well as domain walls, two-dimensional defects, had to be
highly constrained so as to be consistent with
observations~\cite{Zeldovich75,Raphaeli83}.  However, before
observations of the Cosmic Microwave Background (CMB) become available
on a wide range of angular scales, cosmic strings (unidimensional
defects) were weakly constrained by observations.  Moreover, it was
proposed that they could produce the primordial perturbations needed
in the early Universe to form the large-scale structures we observe
today.  In this scenario, cosmic strings seed the formation of
structures by gravitational clustering~\cite{Zeldovich80,Vilenkin81},
achieving a similar result as inflation as far as structure formation
is concerned.

But if inflation and the existence of topological defects without
  inflation are both compelling mechanisms to explain structure
  formation, they lead to dramatically different predictions of the
  CMB anisotropies and polarization angular power spectra:  inflation
  predicts the existence of multiple peaks in all of these
  spectra~\cite{Hu95,Zaldarriaga97}, whereas cosmic defects always
  lead to spectra with a single 
  bump~\cite{Pen97,Battye99,Contaldi99,Durrer99,Pogosian03}.

Therefore, as more and more precise measurements of the CMB were made,
it became possible to test the cosmic strings scenario against
inflation.  These observations now clearly show multiple peaks in the
angular power spectrum of the CMB
anisotropies~\cite{Hinshaw06,Grainge03,Ruhl03,Abroe04,Kuo04,Readhead04},
as well as in the corresponding polarized
signal~\cite{Page06,ReadheadPol04,Leitch05,Barkats05,Montroy05}.
Cosmic defects alone are therefore unable to fit the CMB temperature
angular power spectrum or to give a good description of the
observed~polarization.

But hybrid inflation~\cite{Linde91,Linde94} and supersymmetric hybrid
  inflation~\cite{Copeland94, Dvali94, Linde97}, which predict the
  coexistence of topological defects with an inflationary phase in
  their standard version~\cite{Linde94,Kallosh03}, have been shown to
  be consistent with CMB observations as soon as the first observation
  of a succession of acoustic peaks was made in the CMB temperature
  anisotropies power spectrum~\cite{Bouchet02}.  Later studies have
  explored both cosmic strings and various kind of cosmologically
  acceptable defects in light of recent CMB
  measurements~\cite{Bevis04,Wu05,Fraisse05,Wyman05}.

In this paper, we study what the newly released WMAP results can teach
  us about hybrid inflation\footnote{Unless otherwise stated, we study
  the case where no \emph{a priori} theoretical constraints are taken
  into account.  For example, we do \emph{not} include the fact that
  the spectral index $n_s$ can usually not be smaller than $0.98$.
  This allows our constraints to stay valid even if $n_s \leq 0.98$,
  which can be the case when the radiative corrections arising from
  the introduction of a non-minimal K\"{a}hler potential are taken
  into account.  These two points are discussed, for example,
  in~\cite{BasteroGil06}.}.  We first look at whether introducing
  topological defects to explain WMAP observations is justified in a
  Bayesian analysis.  If it is indeed always possible to add a
  contribution of defects and fit it to the data, this does not tell
  us anything about whether adding this extra parameter gives a better
  description of the observations, and this is what we should first
  check, using the Bayesian evidence as a selection
  tool~\cite{Jeffreys61,MacKay03}.  We then constrain hybrid models
  involving various kind of defects using a full Markov Chain Monte
  Carlo analysis.  Finally, we translate these constraints into bounds
  on the free parameters of D- and F-term inflationary models.


\section{Model selection}

To constrain the values of cosmological parameters, one must first
choose a set of parameters that will then be fitted to the data.  To
do so, a solution is to choose a model ad hoc, with the criteria that
it must both be theoretically attractive and give a good description
of the data.  It turns out that this selection has no unique answer.
It is, for example, possible to fit WMAP data using a running or a
non-running spectral index, both giving a comparable description of
the CMB anisotropies~\cite{Spergel06}.  Therefore, one needs an
additional tool to be able to discriminate between two models of
different dimensions.

\emph{Model selection} makes such a comparison possible.  This
technique rests on the observation that adding a parameter to a given
model and fitting this new model to the data almost always gives a
better fit unless a strong prior is added on the value of this extra
parameter~\cite{ALiddle}.  In other words, a model should be penalized
compared to another one if it requires more parameters to get a
marginally better fit, as should a model requiring fewer parameters
but leading to a less satisfying fit.

In the context of Bayesian analysis, one of the best way of
\emph{grading} a model is to look at its Bayesian
evidence~\cite{Beltran}

\begin{equation}
E = \int{}{}
\mathcal{L}(\bi{s})\,\mathrm{Pr}(\bi{s})\,\mathrm{d}\bi{s}
\end{equation}

\noindent where $\bi{s}$ is a given set of parameters,
$\mathcal{L}(\bi{s})$ the likelihood of this model in light of the
data and $\mathrm{Pr}(\bi{s})$ the priors on the chosen parameters.  A
model is favoured by the data compared to another one if its evidence
is higher.  However, for this to be significant, the difference
$\Delta\,\ln E$ between two models should be at least of order 2.3, in
which case the model of higher evidence is about ten times more likely
to be a better description of the observations than the other
one~\cite{Jeffreys61}.  But we would need a difference of 5 to 6 to
claim that the data decisively reject one of the models compared to
the other one.

We consider two models:  a power-law flat $\Lambda$CDM mo\-del,
characterized by six parameters, and a hybrid model in which the power
spectrum of the CMB anisotropies~is

\begin{equation}
C_{\ell} = (1-\alpha)\,C_{\ell}^{\Lambda \mathrm{CDM}} + 
\alpha\,C_{\ell}^{\mathrm{TD}}
\end{equation}

\noindent where $\alpha$, the contribution of topological defects
(TD), is a seventh parameter and we compute the evidence for each
model by thermodynamic integration.  In practice, this means that $E$
is calculated for each model by running Markov chains in which the
acceptance criterion is given by the likelihood to the power $\theta$
(instead of 1), where $\theta$ is a parameter slowly varying from 1 to
0 as the chain runs.  Following~\cite{Beltran}, we let $\theta$
continuously vary with the number of steps $n$ in the Markov chain.
At the $n\mathrm{th}$~step, $\theta_n=(1-\xi)^n$, where $\xi$ is a
constant empirically set to $5 \times 10^{-5}$ to optimize the speed
and accuracy of the determination of $E$.  The chain is stopped after
$n_\mathrm{max}$ steps, where the step $n_\mathrm{max}+1$ would change
$\ln E$, defined~by 

\begin{equation}
\ln E = \sum_{n=0}^{n_\mathrm{max}}\,
\left[\ln \mathcal{L}(\bi{s})\right]_n\,\xi\,(1-\xi)^n
\end{equation}

\noindent by less than $10^{-3}$.  As we did in~\cite{Fraisse05}, we
generate~the adiabatic spectrum at each step of a Markov chain 
with~\mbox{CMBwarp}~\cite{CMBwarp}.  The results are similar to what
we get with CAMB in the range of multipoles and cosmological
parameters we consider.  The situation is much simpler for topological
defects, as only the overall normalization of the angular power
spectra they induce can change between two consecutive steps.
Therefore, we only need to generate the corresponding spectra once,
before running the chains, and choose an initial normalization.

As it is interesting to be able to compare how the constraints have
evolved between the two WMAP releases, we choose the normalization
used in~\cite{Fraisse05}, for which the power in the angular power
spectrum is kept to the value predicted by the \emph{first-year} (and
not the three-year) WMAP best fit with six parameters, independently
of the value taken by $\alpha$.  The value of the latter is
consequently dependent on this normalization, which one should take
into account when comparing different works.  We then choose the
shapes of the spectra induced by topological defects and their
relative amplitudes in~\cite{Wu05}, which considers all cosmologically
motivated defects along with the differences in the predictions of
different models of the same kind of defects.

We explore the parameter space using flat priors on all our
cosmological parameters such that $(\omega_b,\omega_m) \in [0,1]^2$,  
$h \in [0.5,1.5]$, $A \in [0.5,2.5]$, $\tau \in [0,0.3]$,  
$\alpha \in [0,1]$ and $n_s \in [0,2]$.  We want to get an evaluation
of $\ln E$ to better than 1\%, which requires us to run at least 25
different Markov chains satisfying the stopping criterion for
each~model.

We find that it is statistically impossible to distinguish between a
flat power-law $\Lambda$CDM model and a hybrid model from the point of
view of model selection.  The ratio between the evidence of the
adiabatic model and the evidence of all the other models we tested is
indeed always less than $\mathrm{e}^{0.1 \pm 0.5}$.

In other words, a $\Lambda$CDM model, which is a particular case of
  any hybrid model, is as good a description of the data as a hybrid
  model with topological defects, and there is no obvious need to
  introduce the latter.  The data being indifferent to the set of
  parameters used, the next logical step is to determine what values
  of the cosmological parameters of a hybrid model can fit the data,
  and what are the constraints on the added contribution of
  topological defects.


\section{Best fit hybrid models}

The method used to find the best fit hybrid models is entirely similar
to the one developed in the previous section, except that the
acceptance criterion is governed by the likelihood of the model,
instead of a varying power of the latter.  This analysis therefore
follows the algorithm used in~\cite{Verde} for a flat power-law 
$\Lambda$CDM model.

It is also much less computationally expensive than the model
selection analysis.  Performing the tests described
in~\cite{Fraisse05}, we indeed find that running eight chains of 
$2 \times 10^5$ steps each is sufficient to get a value of the Gelman
and Rubin convergence diagnostic down to 1.1~\cite{Gelman}.  Getting
the best fit models is therefore about four times faster than checking
if the data favours the addition of an extra parameter.

\begin{table}
\caption{\label{Bounds}Upper limits on the contribution of defects to
  the CMB angular power spectrum (\%), and corresponding strings'
  tensions ($G\mu$) for three different models.  The tension given for
  global strings is the one a local string with the same spectrum
  would have.  In a given cell, the first (second) number is the 68\%
  (95\%) confidence level upper~bound.}
\lineup
\begin{indented}
\item[]\begin{tabular}{@{}llll}
\br
Defects&&Upper bound&$G\mu \times 10^7$\\
\mr
Global strings&\cite{Pen97}&13\% -- 18\%&2.4 -- 2.8\\

Local strings&\cite{Pogosian03}&{\0}7\% -- 11\%&2.1 -- 2.6\\

Local strings&\cite{Battye99}&{\0}5\% -- {\0}7\%&2.1 -- 2.5\\
\br
\end{tabular}
\end{indented}
\end{table}

We give the 68\% and 95\% confidence level upper bounds on the
contribution of one model of global cosmic strings and two models of
local cosmic strings in \tref{Bounds}.  In addition to these models,
we also considered the model of~\cite{Contaldi99}, for which we get
similar results.  Finally, global and non-topological textures produce
spectra indistinguishable from the model of global strings we
use~\cite{Pen97}, so that their contributions are subject to the same
upper~bound.

We find that the results are \emph{weakly} dependent on the model of
defects we consider.  Even when we use the spectrum induced by global
strings to get a constraint on $G\mu$, the corresponding bound differs
from the value derived with local strings by only 8\% at 
$3\,\sigma$.  In each case, the best fit deviates from 0\% at
$1\,\sigma$, as one would expect when an extra parameter is added.

Combining these different results, we can choose a conservative upper
bound (68\%~CL) of $G\mu \leq 2.1 \times 10^{-7}$, corresponding to a
contribution of 6\% of local strings to the CMB temperature angular
power spectrum.  This is all the more reasonable that recent
measurements of the third acoustic peak (weakly constrained by the new
WMAP results~\cite{Hinshaw06}) by VSA~\cite{Grainge03},
BOOMERanG~\cite{Ruhl03}, and CBI~\cite{Readhead04} suggest that it
lies close to the predicted $\Lambda$CDM spectrum~\cite{Spergel06}.
Therefore, as defects produce less power at the scale of the third
acoustic peak than at the one corresponding to the second
peak~\cite{Wu05}, the contribution of defects should go down when
these data are taken into account.


\section{Constraints on hybrid inflationary models}

Standard hybrid inflation can avoid the formation of various kind of
defects but cosmic strings are particularly
unavoidable~\cite{Jeannerot97,JeannerotRocher} in the context of SUSY
GUTs\footnote{Note that this is not the case for all kinds of hybrid
  inflationary models, as some of them, e.g. smooth and shifted hybrid
  inflations, do \emph{not} lead to the formation of
  defects~\cite{JeannerotPostma}.}.  A constraint on the existence of
cosmic strings is therefore a direct cosmological constraint on these
unification theories.

GUT strings formed at the end of F-term inflation do not saturate the
 Bogomolny bound, whereas strings formed in D-term inflation
 do~\cite{JeannerotPostmaJHEP}.  The energy scale of the SSB leading
 to the production of the corresponding defects is therefore dependent
 on what scenario one considers.  However, this can be understood with
 a unified formalism in which the mass scale $M$ of a theory is given
 by

\begin{equation}
\label{eqn:mass}
M = M_{\mathrm{Pl}}\,
\sqrt{\frac{G \mu}{2 \pi}}\,\frac{1}{\sqrt{\epsilon}}
\end{equation}

\noindent where $M_{\mathrm{Pl}}$ is the Planck mass and $\epsilon$ is
a dimensionless parameter whose value is in $[0,1]$ and depends on the
type of inflation one considers.  Using the constraint on $G\mu$ that
we previously derived, it is then possible to find an upper bound on
$M$ as a function of $\epsilon$ for the theories mentioned above.

\begin{figure}[h]
\caption{\label{fig:M-k}Constraints on the $(M,\kappa)$-space of
  F-term hybrid inflation from one and three years of WMAP results.
  The blue (respectively orange) curve corresponds to the 68\%
  (resp. 95\%) confidence limit (CL) upper bound using three years of
  WMAP data.  The blue-shaded area is the region of the parameter
  space allowed at the 68\%~CL.  The orange-shaded area is the
  additional allowed region if one considers the 95\%~CL.  The green
  curve corresponds to the average 95\%~CL upper bound using one year
  of WMAP data as derived from~\cite{Bevis04,Wu05,Fraisse05}.  The
  green-shaded area is therefore the region of parameter space allowed
  by one year, but excluded by three years, of WMAP data. 
  }
\begin{indented}
\item[]
\includegraphics[scale=0.5]{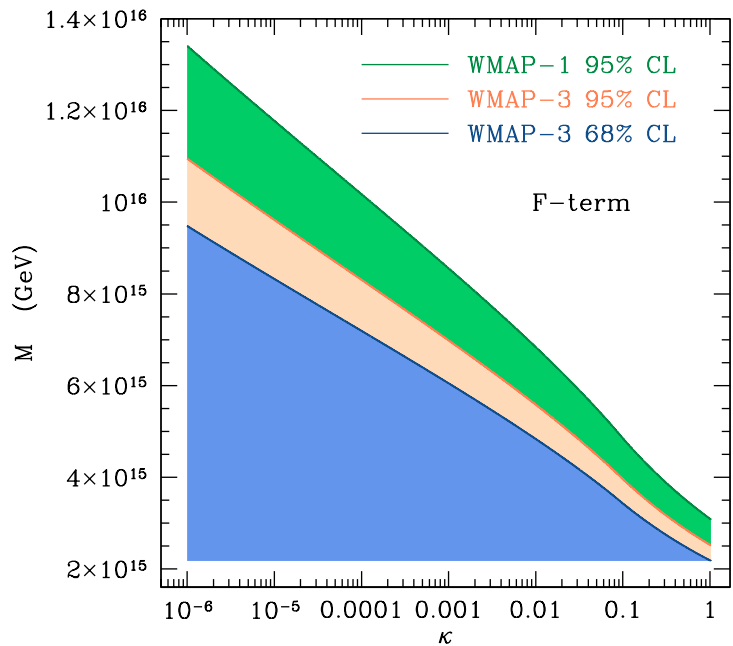}
\end{indented}
\end{figure}

In the case of D-term inflation, in which strings satisfy the
Bogomolny limit, $\epsilon = 1$.  In GUT F-term inflation, the SUSY
superpotential $W_\mathrm{F}$ can be written as

\begin{equation}
W_\mathrm{F} = \kappa\,S\left(\phi_ + \phi_- - M^2\right)
\end{equation}

\noindent where $S$ is the slow-rolling field of the theory, $\phi_+$
and $\phi_-$ are chiral GUT Higgs superfields, and $\kappa$ is a
coupling constant.  $\epsilon$ is then a function of $\kappa$ whose
variation obeys a different law for $\kappa \leq 10^{-1}$ and 
$\kappa \geq 10^{-1}$.  In the latter case, $\epsilon$ is well
approximated by 
$\epsilon = 1.04 \times \kappa^{0.39}$~\cite{JeannerotPostmaJHEP}.
Moreover, when $\kappa \to 0$,
$\epsilon \to 2.4/\ln(2/\kappa^2)$~\cite{HillHodgesTurner}.
Interpolating between those two regimes, one can thus take $\epsilon$
as 

\begin{equation}
\epsilon =\left\lbrace
\begin{array}{lrl}
1.04 \times \kappa^{0.39} & \mathrm{if} & \kappa \geq 10^{-1}\\
2.4 \, \left[\ln(2/\kappa^2)\right]^{-1} - 2.93 \times 10^{-2} &
\mathrm{if} & \kappa \leq 10^{-1}\\
\end{array}
\right.
\end{equation}
 
\noindent where this last constant is chosen so that $\epsilon$ be a
continuous function of $\kappa$.  Due to this dependence of $\epsilon$
upon $\kappa$ in GUT F-term inflation, our limits on $G\mu$ translate
into a constraint on what area of the $(M,\kappa)$-space is allowed by
three years of WMAP data.  These results are shown in
figure~\ref{fig:M-k}.

In the case of D-term inflation, as $\epsilon = 1$, it is possible to
get a direct constraint on the mass scale of the theory, using 
equation~(\ref{eqn:mass}).  Moreover, the SUSY superpotential
$W_\mathrm{D}$ of the theory can be written 

\begin{equation}
W_\mathrm{D} = \lambda\,S\,\phi_+ \phi_-
\end{equation}

\noindent where we use notations similar to the ones used for
$W_\mathrm{F}$, and $\lambda$ is the superpotential coupling constant.
For this kind of hybrid inflationary model, \cite{Rocher} showed that
it is possible to find a relationship between $\lambda$ and the
strings' contribution to the CMB angular power spectrum, as well as
between $\lambda$ and the Fayet-Iliopoulos term~$\xi$, which are two
of the three free parameters of D-term inflation.  However, the
relationship between $\lambda$ and the strings' contribution is
dependent on the third free parameter of the theory, namely the gauge
coupling constant $g$.  Though the effect of $g$ on the shape of this
function is shown to be pretty dramatic in~\cite{Rocher}, it is hard
to set a stringent constraint on the gauge coupling constant from a
given contribution of D-term strings, unless the latter is very low (a
few hundredth of a percent) or very high (close to 100\%).  From our
constraints, we can only feel confident that the maximum value of $g$
allowed by three years of WMAP data is of the order of $10^{-2}$,
leading to

\begin{equation}
\lambda \leq 2.3 \times 10^{-5} \qquad \mathrm{and} \qquad
\sqrt{\xi} \leq 2.0 \times 10^{15}\,\mathrm{GeV}.
\end{equation}

\noindent This last parameter can also be considered as the
inflationary scale of the theory.  We summarize the constraints
derived in this section for D-term inflation and for one and three
years of WMAP observations in \tref{SUSY}.

\begin{table}[h]
\caption{\label{SUSY}Upper bounds on the D-term strings
  ($M_\mathrm{s}$) mass scale and the D-term coupling constant
  ($\lambda$) and inflationary scale ($\sqrt{\xi}$) from one and three
  years of WMAP observations.}
\lineup
\begin{indented}
\item[]\begin{tabular}{@{}lll}
\br
Parameters&1-year&3-year\\
\mr

$M_s \times 10^{-15}$ (GeV)&$2.9$&$2.2$\\

$\lambda \times 10^5$&$4.5$&$2.3$\\

$\sqrt{\xi} \times 10^{-15}$~(GeV)&$2.6$&$2.0$\\
\br
\end{tabular}
\end{indented}
\end{table}


\section{Conclusions}

Using three years of WMAP data, we showed that it is not possible to
exclude hybrid inflationary models by a statistical analysis.  The
data does not prefer a $\Lambda$CDM description to a model involving
global strings, global textures or local strings, even if we take into
account the dimensions of the models we compare, as was the goal of
our model selection analysis.  On the other hand, there is no
compelling evidence requiring the introduction of cosmic defects in
addition to the six parameters of a power-law flat $\Lambda$CDM model.
Moreover, the WMAP three-year data is precise enough to set stronger
constraints on topological defects than what was previously found with
one year of WMAP observations.

We find that the tension of cosmic strings must be less than 
$G\mu = 2.6 \times 10^{-7}$ at $3\,\sigma$, a more conservative value
being $G\mu \leq 2.1 \times 10^{-7}$, which corresponds to the 68\%
confidence level upper bound\footnote{After this paper was submitted,
  \cite{Seljak06} reported similar constraints, using a combination of
  WMAP three-year results and Lyman-alpha forest, galaxy clustering
  and supernovae constraints.}.  This limit does \emph{not} depend on
the model of local strings we use.  Even when we consider that local
strings are similar to global strings, the discrepancy is only of
order 8\% at $3\,\sigma$.  From these constraints, we then derive
upper bounds on the free parameters of D- and F-term inflation, a
summary of which is given in figure~\ref{fig:M-k} and \tref{SUSY}.

Hybrid inflation is therefore becoming more and more constrained by
CMB measurements, but not yet to a level requiring the introduction of
un\-natural mo\-di\-fi\-cations in the theory.  As this scenario does
not involve more fine tuning than \emph{classical} inflation, is
perfectly consistent with all current observations, and is naturally
predicted by SUSY GUTs, it is worth keeping it in mind.


\ack

We thank David Spergel for his support throughout the course of this
work, Juan Maldacena for illuminating discussions on SUSY GUTs, Licia
Verde for her help in the use of CMBwarp and of the WMAP likelihood
code, Joanna Dunkley for comments on a draft version of this paper,
and the anonymous referee for comments that helped to improve the
accuracy and clarity of the material in this paper.  This work was
supported by NASA Astrophysics Theory Program grant NNG04GK55G and
Princeton University, and made use of computational facilities
supported by NSF grant AST-0216105.  We acknowledge the use of the
Legacy Archive for Microwave Background Analysis (LAMBDA).  Support
for LAMBDA is provided by the NASA Office of Space Science.  This
research made use of NASA's Astrophysics Data System
Bibliographic~Services, and some of the results in this paper have
been derived using the HEALPix~\cite{Healpix}, CAMB~\cite{Lewis99} and
CosmoMC~\cite{CosmoMC} packages. 


\end{document}